\documentclass{ws-p8-50x6-00}%
\usepackage{graphics,pennames}

\newcommand{\Zvgv}{{\mathrm Z^*} / \gamma^*}

\begin{document}

\title{Constraining CP-Violating TGCs and Measuring W-Polarization at OPAL}

\author{Isabel Trigger}

\address{Division EP, CERN, CH-1211 Gen\`eve-23, Switzerland\\E-mail:
  Isabel.Trigger@cern.ch} 

\maketitle
{ \vspace{-5cm}
  \flushright \parbox{\textwidth}{
  \bf \flushright
  OPAL Conference Report CR456
  \\ \today } 
}

\vspace{4cm}
\abstracts{
A measurement of the polarization of $\mathrm{W}$ bosons in
semi-leptonically decaying $\mathrm{W}$ pairs produced at 189~GeV is
presented.  Rates of longitudinally and transversely polarized $\mathrm{W}$
bosons and correlation between two $\mathrm{W}$ bosons are studied.  The spin
properties of the leptonically decaying $\mathrm{W}$ boson in the $\mathrm{W}$ pairs was
used to measure the CP-violating trilinear gauge boson couplings.
These results are compared with Standard Model expectations.}

\section{The Spin Density Matrix for \PWp\PWm}

Polarization properties of \PWp\PWm{} produced in \Pep\Pem{}
collisions are summarized by the two-particle joint spin density
matrix (SDM).  The SDM is the product of the amplitudes for
producing a \PWp{} and \PWm{} of respective helicities $\tau_+$, 
$\tau_-$:
\begin{equation}
  \rho_{\tau_{-}{\tau^{\prime}}\!\!_{-}\tau_{+}{\tau^{\prime}}\!\!_{+}}(s,\cos\theta_{\rm W}) = 
\frac{\sum_{\lambda}F^{(\lambda)}_{\tau_{-}\tau_{+}}(F^{(\lambda)}_{\tau_{-}^{
\prime}\tau_{+}^{\prime}})^{*}}{\sum_{\lambda\tau_{+}\tau_{-}}|F^{(\lambda)}_{
\tau_{-}\tau_{+}}|^{2}} \, .
\label{eq:jsdm}
\end{equation}
The diagonal elements are probabilities of producing
\PWp\PWm{} with the corresponding helicity combinations, and are
strictly real.  The potentially complex off-diagonal terms represent
interference between helicity states.
The data considered in the analysis were collected with the
OPAL detector in 1998, at a centre-of-mass energy of 189~GeV.  They
correspond to an integrated luminosity of $\sim 183\mbox{ pb}^{-1}$, with
1065 events identified as $\PWp\PWm\to \mathrm{q {\bar{q}}}
\ell \nu$.
Only semi-leptonic decays are used for this analysis,
as fully leptonic and fully hadronic decays cannot be unambiguously
reconstructed. 
Due to the restricted sample size, it is not possible to measure all
81 elements of the joint SDM; however, all elements of the
nine-element single-particle SDM, obtained by summing over the
helicity states of the \PWp, may be measured: 
\begin{equation}
  \rho^{W^{-}}_{\tau_{-}{\tau^{\prime}}\!\!_{-}}(s,\cos\theta_{\rm W}) = \sum_{\tau_{+}}\rho_{\tau_{
-}{\tau^{\prime}}\!\!_{-}\tau_{+}\tau_{+}}(s,\cos\theta_{\rm W}) \, .
\label{eq:ssdm}
\end{equation}
Its diagonal elements are the probabilities of producing a
\PWm{} with helicity $+1, 0$ or $-1$. In order to use
both $\mathrm{ q {\bar q}}\ell^-\bar{\nu}$ and $\mathrm{ q {\bar q}}\ell^+\nu$,
CPT is assumed.

The five kinematic variables used in the analysis are the cosine of
the polar production angle of the \PWm{}
in the laboratory frame ($\cos\theta_W$), and the
decay angles given by the directions of the fermion and anti-fermion
with respect 
to the \PWp{} and \PWm{} respectively in the $\mathrm{W}$ rest-frames  
($\cos\theta_\ell^\ast,\phi_\ell^\ast,$
$\cos\theta_j^\ast, \phi_j^\ast)$, with an
ambiguity of $\pi$ for both jet angles due to the difficulty of
distinguishing the up-quark jet from the anti--down-quark jet.
With the hadronic part of the event, it is therefore only
possible to measure combinations of SDM elements which are symmetric
under $\cos\theta^\ast \to -\cos\theta^\ast$ and $\phi^\ast \to
\phi^\ast+\pi$.
Fortunately, these include $\rho_{++}+\rho_{--}$ and $\rho_{00}$, so
the full two-particle SDM element
$\rho_{0000}$ and the combinations
$\rho_{++++}+\rho_{++--}+\rho_{--++}+\rho_{----}$ and
$\rho_{++00}+\rho_{--00}+\rho_{00++}+\rho_{00--}$ may be measured.

SDM elements are measured by forming histograms of the
$\cos\theta_W$ distribution obtained in the data, and weighting each event
by a projection operator which is a function of
$\cos\theta_\ell^\ast,\phi_\ell^\ast,\cos\theta_j^\ast,
\phi_j^\ast$.  Different operators project out each independent
element of the SDM.  
\begin{figure}[htbp]
\begin{center}
\resizebox{1.\textwidth}{!}{\includegraphics{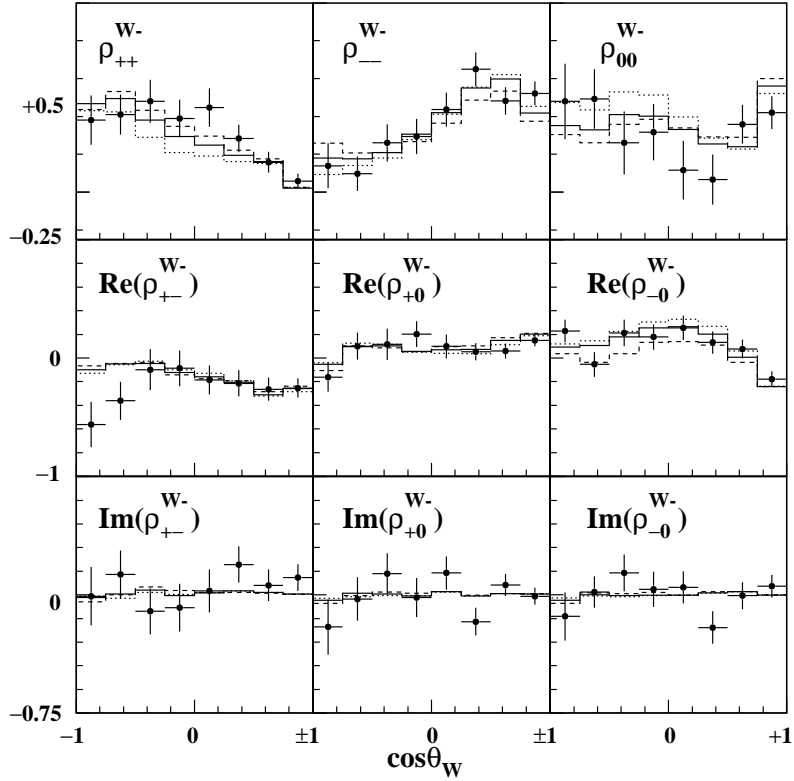}}
\end{center}
  \caption{SDM elements from
    leptonically decaying W bosons in $qq\ell\nu$ data events.
    Points are OPAL data, with statistical and systematic
    errors. Histograms show
    Monte Carlo predictions with full detector
    simulation. The solid line shows the SM
    expectation and the dotted (dashed) line that for $\Delta g^{\rm
    z}_{1}$ = +0.5 ($-$0.5).  
    \label{fig:sdm}}
\end{figure}
Results are shown in figure~\ref{fig:sdm}.

Constraints arise from imposing tree-level CPT-invariance:
\begin{eqnarray}
{\rm Re}(\rho^{W^{-}}_{\tau_{1}\tau_{2}}) &-&{\rm
  Re}(\rho^{W^{+}}_{-\tau_{1}-\tau_{2}}) = 0 \label{eq:cptreal}\\
{\rm Im}(\rho^{W^{-}}_{\tau_{1}\tau_{2}}) &+& {\rm
  Im}(\rho^{W^{+}}_{-\tau_{1}-\tau_{2}}) = 0 \, . \label{eq:cptim}
\end{eqnarray}
Since CPT has already been assumed, any deviation from these must
arise from loop effects.
CP-invariance would further imply that:
\begin{equation}
  {\rm Im}(\rho^{W^{-}}_{\tau_{1}\tau_{2}}) - {\rm
  Im}(\rho^{W^{+}}_{-\tau_{1}-\tau_{2}}) = 0 \, . \label{eq:cpim}
\end{equation}
Combining equations~\ref{eq:cptim} and~\ref{eq:cpim} shows that all
single-particle SDM coefficients must be strictly real in the absence
of CP-violation.

\section{Helicity Cross-Sections} \label{sec:polar}
  Differential production cross-sections for $\mathrm{W}^\pm$ of
particular helicities are:
\begin{equation}
  \frac{{\rm d}\sigma_h}{{\rm d}\!\cos\theta_{\rm W}} = f_h 
\frac{{\rm d}\sigma}{{\rm d}\!\cos\theta_{\rm W}} \, , \label{eq:crsec}
\end{equation}
where $f_{\rm T}=\rho_{++}+\rho_{--}$, $f_{\rm L}=\rho_{00}$, $f_{\rm
  TT}=\rho_{++++}+\rho_{++--}+\rho_{--++}+\rho_{----}$, $f_{\rm
  LL}=\rho_{0000}$, $f_{\rm
  TL}=\rho_{++00}+\rho_{--00}+\rho_{00++}+\rho_{00--}$.
Corresponding total cross-sections may be obtained by integrating over
the full range of ${\rm d}\!\cos\theta_{\rm W}$.  Results are
given in table~\ref{tab:pol}.  Correction factors are included for
detector and reconstruction effects.
\begin{table}[htbp]
\begin{center}
\begin{tabular}{|l|c|c|} \hline
  & {Data} & {SM Exp.} \\
 \hline
\underline{$\sigma_{\rm T}/\sigma_{\rm total}$} & &   \\
W$ \rightarrow \ell{\nu}$ & 0.842 $\pm$ 0.048 $\pm$ 0.023 & 0.746
 $\pm$ 0.006\\  
W$\rightarrow {\rm q}{\rm q}$ & 0.738 $\pm$ 0.045 $\pm$ 0.025 & 0.741
 $\pm$  0.006\\  
All  & 0.790 $\pm$ 0.033 $\pm$ 0.016 & 0.743 $\pm$ 0.004\\
\hline 
\underline{$\sigma_{\rm L}/\sigma_{\rm total}$} & &  \\
W$\rightarrow \ell{\nu}$ & 0.158 $\pm$ 0.048 $\pm$ 0.023  & 0.254
 $\pm$ 0.006 \\  
W$\rightarrow {\rm q}{\rm q}$ & 0.262 $\pm$ 0.045 $\pm$ 0.025 & 0.259
 $\pm$ 0.006\\   
All  & 0.210 $\pm$ 0.033 $\pm$ 0.016 & 0.257 $\pm$ 0.004\\\hline 
\end{tabular}  
\end{center}
\begin{center}
\begin{tabular}{|l|l|l|} \hline
 &  Measured  & Expected\\
 \hline 
$\sigma_{\rm TT}/\sigma_{\rm total}$ & 0.781 $\pm$ 0.090 $\pm$ 0.033 &
 0.572 $\pm$ 0.010 \\\hline 
{$\sigma_{\rm LL}/\sigma_{\rm total}$} & {0.201 $\pm$ 0.072 $\pm$ 0.018}  & {0.086 $\pm$ 0.008} \\\hline
$\sigma_{\rm TL}/\sigma_{\rm total}$ & 0.018 $\pm$ 0.147 $\pm$ 0.038 &
 0.342 $\pm$ 0.016 \\\hline 
\end{tabular}
\end{center}
\caption{Fractions of $\mathrm{W}$ polarizations, and of \PWp\PWm{}
  pairs of each helicity combination. Expected values are from
 generator level EXCALIBUR Monte Carlo.
 \label{tab:pol}} 
\end{table}
The \PWp{} and \PWm{} polarizations are about 7\% correlated. 
This effect is included in the systematic errors.  Total
cross-sections for the various helicity states are very strongly
correlated.
It should be noted that $\rho_{0000}$ can have no CP-violating
contributions,\cite{rho0000}
and $\frac{{\rm d}\sigma_{\rm LL}}{{\rm
d}\!\cos\theta_{\rm W}}$ is thus completely insensitive to
CP-violation.  The helicity fractions for TT, LL and TL differ by
about $2\sigma$ from Standard Model (SM) predictions,
giving a $\chi^2$ probability of about 10\% for SM
compatibility.

\section{Triple Gauge Boson Couplings} \label{sec:tgc}
Of nine possible \PWp\PWm{} helicity pairings, seven are allowed
in the $s$-channel processes $\Pep\Pem\to \Zvgv \to \PWp\PWm$
($+-$, $-+$ occur only in $t$-channel $\nu$-exchange, as $|\tau_+
- \tau_-|=2$).  There are then seven 
free parameters in the Lagrangian describing each of the triple
gauge boson couplings $\PWp\PWm\PZz$ and $\PWp\PWm\Pgg$,
here called $\kappa_V, g_1^V, \lambda_V, g_5^V, \tilde\kappa_V,
g_4^V, \tilde\lambda_V$ ($V=\PZz, \Pgg$).  The first two are
unity in the SM, and the others zero.  The first three
conserve C and P, $g_5^V$ violates C and P but conserves CP, and the
last three violate CP.

\begin{figure}[htbp]
\begin{center}
\resizebox{1.\textwidth}{!}{\includegraphics{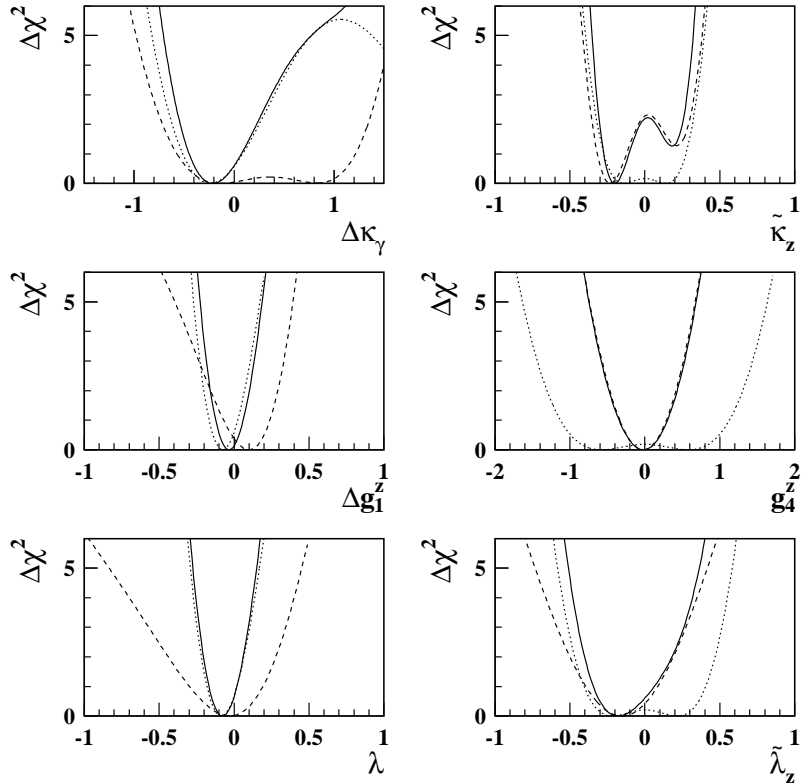}}
\end{center}
  \caption{The $\chi^2$ plots for the fits to the CP-conserving and
    CP-violating anomalous couplings.
    The dotted line is the fit to just the $\cos\theta_{\rm
    W}$ distribution. The solid line is the combined fit. All fits
    include systematic uncertainties.\label{fig:chi}} 
\end{figure}

The real elements of the SDM are sensitive to all of the coupling
parameters; the imaginary elements are also sensitive to
CP-violating parameters.  It is impossible to fit all fourteen
parameters simultaneously with the limited data sample.  Each
is measured separately, with the others set to their
SM values, except those related to the tested parameter
through $SU(2)\times U(1)$ symmetry\cite{goun} ($\tilde\kappa_Z =
-\tan^2\theta_W \tilde\kappa_\gamma$; $\tilde\lambda_Z =
\tilde\lambda_\gamma$; $g_4^Z = g_4^\gamma$).
This leaves three independent parameters to test, 
chosen to be $\tilde\kappa_Z, \tilde\lambda_Z, g_4^Z$.
There are strong constraints on CP-violation in
electromagnetic interactions from the constraints on the electric
dipole moment of the neutron,\cite{neutron} which is why an
alternative would be to ignore the gauge symmetry constraints and set
\PWp\PWm\Pgg{} couplings to zero.  Coupling values from a fit to
the SDM elements are given in table~\ref{tab:tgc}.
\begin{table}[htbp]
\begin{center}
\begin{tabular}{|l|c|c|c|} \hline
Fit & $\tilde{\kappa}_{\rm z}$ & $g^{\rm z}_{4}$ & $\tilde{\lambda}_{\rm z}$ \\\hline
SDM Elements & $-0.19^{+0.08}_{-0.07}$  & $\;\;\;$$0.00^{+0.21}_{-0.20}$ & $-0.12^{+0.17}_{-0.16}$  \\
$\cos\theta_{\rm W}$ & $-0.19^{+0.46}_{-0.08}$ & $\;\;\;$$0.7^{+0.4}_{-1.8}$ & $-0.29^{+0.69}_{-0.11}$  \\\hline
Combined & $-0.19^{+0.06}_{-0.05}$ & $\;\;\;$$0.01^{+0.22}_{-0.22}$ & $-0.19^{+0.18}_{-0.13}$  \\
Expected Stat. Error  & $\;\;\;\pm$0.11 & $\;\;\;\pm$0.19 & $\;\;\;\pm$0.12   \\\hline
Final Fit  & & & \\
Including Systematics   &
\raisebox{1.5ex}[0pt]{$-0.20^{+0.10}_{-0.07}$} &
\raisebox{1.5ex}[0pt]{$-0.02^{+0.32}_{-0.33}$} &
\raisebox{1.5ex}[0pt]{$-0.18^{+0.24}_{-0.16}$} \\\hline 
\end{tabular}
\caption{Measured values of CP-violating TGC parameters. Both the SDM
  elements
  and the $\cos\theta_{\rm W}$ production distribution
  are used in the calculation. Errors are statistical only except in
  the case of the final combined fit.} 
\label{tab:tgc}
\end{center}
\end{table}
Table~\ref{tab:tgc} also includes results from a $\chi^2$ fit to the
$\cos\theta_W$ distribution.
The $\cos\theta_W$ distribution is the most
sensitive to variations in CP-conserving couplings, but is
relatively insensitive to CP-violating couplings.
Figure~\ref{fig:chi} shows $\chi^2$ curves for TGCs measured from
SDM elements and from $\cos\theta_W$ distributions.  The
real SDM elements and $\cos\theta_W$ are sensitive only to the
magnitude of CP-violating couplings (their dependence on the
couplings is quadratic), and so have a double minimum.  The
imaginary SDM elements depend linearly on the
CP-violating couplings and can thus lift the degeneracy.
The CP-conserving couplings measured from the SDM elements are fully
compatible with the results of the OPAL optimal observable
analysis.\cite{oo} 

\section{Conclusions}
The SDM method allows direct measurement of the fraction of $\mathrm{W}$ bosons
produced with longitudinal polarization.  This longitudinal component
of the $\mathrm{W}$ is a result of the electroweak symmetry breaking
mechanism.  It also provides constraints on CP-violation in TGCs.
All results are compatible with SM predictions.  This
analysis is more fully described elsewhere.\cite{sdm}

\end{document}